# Selective scattering of blue and red light based on silver and gold nanocubes

Yiyang Ye,[1*] and T. P. Chen,[1*]

[1]*School of Electrical and Electronic Engineering, Nanyang Technological University, Singapore 639798, Singapore*
*E-mail: yeyi0005@e.ntu.edu.sg, echentp@ntu.edu.sg

## Abstract

Selective scattering of red-, green- and blue-light and transmitting other visible light to achieve transparent projection screen has been proposed recently based on metallic nanoparticle's localized surface plasmon resonance (LSPR). However, dielectric ($TiO_2$) substrate/silver (Ag) nanocube structure was only demonstrated to selectively scatter blue-light in the backward direction. Given human eyes are sensitive to green-light, it is of interest to find out how to achieve selective forward-scattering of blue- and red-light and selective backward-scattering of red-light. In this work, through numerical simulation, forward and backward scattering properties of dielectric substrate/Ag (or gold) nanocube structures are investigated. And based on these properties, three designs are proposed which can achieve selective scattering of blue- and red-light in forward or backward or both directions.

## Introduction

Recently, the idea of wavelength-selective scattering of light to achieve transparent projection screen has been proposed based on metallic nanoparticle's localized surface plasmon resonance (LSPR),[1-4] where in the ideal case metallic nanoparticles dispersed in a transparent matrix only selectively scatter red- green- and blue-light and transmit visible light of other colours. Nanospheres and ellipsoids achieve selective scattering of red-, green- or blue-light by taking advantage of the nanoparticle's dipolar resonance peak,[1, 3-4] while Ag nanocubes simultaneously achieve selective scattering of blue- and red-light by a splitting of resonance peak when it is placed on a dielectric substrate.[2] For the transparent projection screen achieved by depositing Ag nanocubes on $TiO_2$ thin film reported in previous work,[2] since it only considers enhancement of back-scattering and its back-scattering of red-light is much weaker than that of blue-light, given human eyes are sensitive to green-light,[5] it is of interest to find out how to achieve selective forward-scattering of blue- and red-light and further enhancement of selective backward-scattering of red-light. In this work, scattering properties of Ag and gold (Au) nanocubes deposited on dielectric ($TiO_2$ or glass) substrate are studied, and three designs for selective scattering of blue and red light are shown, of which, only one is suitable for simultaneous forward and backward selective scattering of blue and red light, the other two are only suitable for either forward or backward selective scattering of blue and red light.

## Simulations and discussion

All simulated results shown in this work are calculated by a numerical simulation method known as Finite-Difference Time-Domain (FDTD).[6] The commercial software to implement FDTD is "FDTD Solutions". To determine the refractive index of $TiO_2$ to be used in the simulation, we deposited a layer of $TiO_2$ thin film with a thickness of 120 nm on a piece of silicon wafer by RF sputtering process followed by calcination at 500 °C for 1 h.  The refractive index averaged in the visible wavelength range (400 nm ~ 800 nm) of the $TiO_2$ thin film is 2.3 as obtained with spectroscopic ellipsometry.[7] For a metallic nanocube placed on dielectric substrate, back-scattered light is detected by a box-shaped monitor placed on the same side (with respect to the metallic nanocube) where light is injected, whose face nearest and parallel to the air/ dielectric interface is open and passes through the geometrical centre of the metallic nanocube, while forward-scattered light is collected by another box-shaped monitor placed on the other side, whose open-face shares the same plane with that of the back-scattered light monitor. In our simulation, the shared open-face of the two box-shaped monitors has a dimension of 400 nm by 400 nm. The source type is "Total-field scattered-field" (TFSF) so that only the scattered light will be detected by monitors.[8] The grid size of the three-dimensional mesh override region is 1 nm. The edge and corner of a metallic nanocube is rounded by a radius which is 10% of the metallic nanocube's edge length, and the dielectric constants of Ag and Au are taken from reference.[9]

The resonance peak splitting of a Ag nanocube when it is placed on a dielectric substrate originates from the substrate-mediated coupling between a bright dipolar and a dark quadrupolar plasmon mode, and the higher the refractive index of the substrate, the larger the peak splitting.[10-15]

From our simulation, it is found that the scattering spectrum of a Ag nanocube is affected not only by the presence of dielectric substrate, but also by propagation direction of incident light when the Ag nanocube is placed on a dielectric substrate, as shown in Figure 1, where scattering cross section is the ratio of the total scattered power to intensity of the incident light. In Figure 1, a splitting of scattering peak is observed when the Ag nanocube in air is placed on $TiO_2$ substrate, and the scattering peak around 440 nm is stronger than the one around 710 nm when light is incident from air to $TiO_2$ substrate (Air-to-$TiO_2$ in short), and the other way around when light is incident from $TiO_2$ substrate to air ($TiO_2$-to-Air in short).

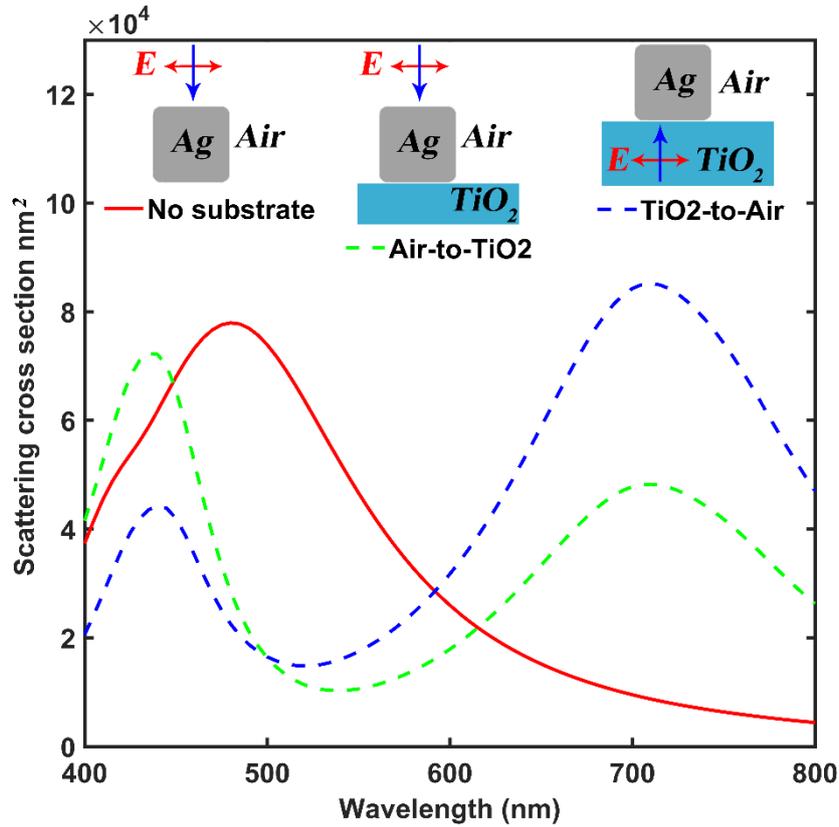

***Figure 1.*** *Scattering cross sections (in nm²) calculated for Ag nanocube with edge length of 100 nm in three setups: without dielectric substrate (solid red line), with TiO₂ substrate and light incident from air to TiO₂ (dashed green line), with TiO₂ substrate and light incident from TiO₂ to air (dashed blue line). Insets illustrate the propagation direction and polarization of light.*

The Air-to-TiO₂ and TiO₂-to-Air scattering spectrums for Ag nanocube placed on TiO₂ substrate as shown in Figure 1 are further decomposed into forward and backward scattering, respectively illustrated in Figure 2 (a) and Figure 3 (a). It is observed that the forward scatterings are quite different from backward scatterings. For the two peaks of the Air-to-TiO₂ scattering as shown in Figure 1, their near electric field amplitude $|E|$ spatial plots are shown in Figure 2 (b) and (c) respectively, while for the two peaks of the TiO₂-to-Air scattering, their $|E|$ spatial plots are shown in Figure 3 (b) and (c) respectively. In Figures 2 and 3, by correlating peaks of forward and backward scattering spectrums to respective near field plots, an interesting finding is described in the following. When the end of nanocube with higher $|E|$, which is the active region of free electrons' resonant oscillation, is nearer to incident light source (as compared to the other end of nanocube with lower $|E|$), backward scattering dominates around the peak wavelength; and in the reverse case, i.e., when the end of nanocube with higher $|E|$ is away from incident light source, forward scattering dominates around the peak wavelength.

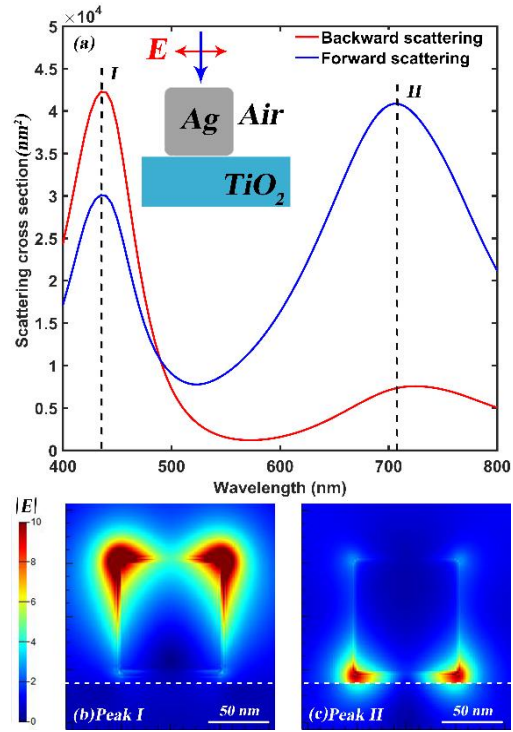

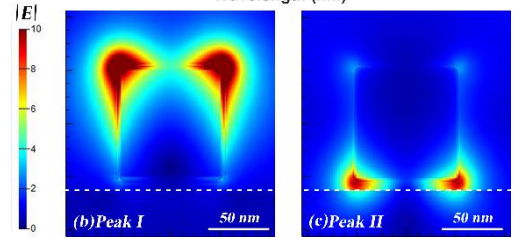

***Figure 2.*** *(a) Forward and backward components of scattering cross section of 100-nm Ag nanocube placed on TiO₂ substrate, with the light propagation direction (Air-to-TiO₂) and polarization illustrated by the inset. (b) Near electric field amplitude |E| spatial plot at Peak I (wavelength =435 nm) as indicated in (a). (c) Near electric field amplitude |E| spatial plot at Peak II (wavelength =710 nm) as indicated in (a). The white dotted line in (b) and (c) indicates the Air/TiO₂ interface. The face monitor for near field recording is parallel to the plane as shown by the inset in (a) and is placed 1 nm away from the front face of Ag nanocube.*

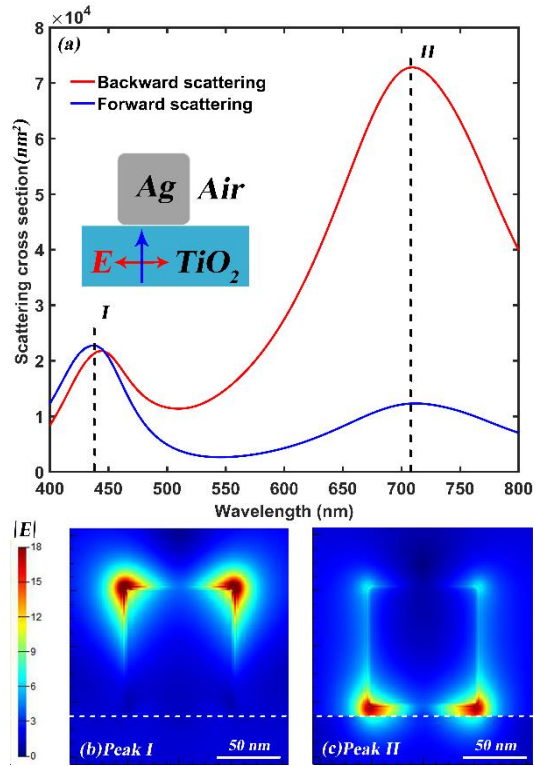

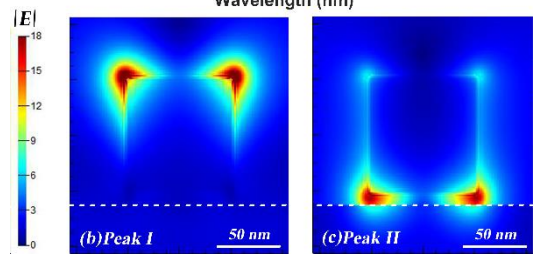

***Figure 3.*** *(a) Forward and backward components of scattering cross section of 100-nm Ag nanocube placed on TiO₂ substrate, with light propagation direction (TiO₂-to-Air) and polarization illustrated by the inset. (b) Near electric field amplitude |E| spatial plot at Peak I (wavelength =440 nm) as indicated in (a). (c) Near electric field amplitude |E| spatial plot at Peak II (wavelength =710 nm) as indicated in (a). The white dotted line in (b) and (c) indicates the Air/TiO₂ interface. The face monitor for near field recording is parallel to the plane as shown by the inset in (a) and is placed 1 nm away from the front face of Ag nanocube.*

Therefore, from Figures 2 and 3, for backward scattering, the Air-to-TiO₂ setup may offer a solution for blue-light (wavelength of about 440 nm to 460 nm) selective scattering while the TiO₂-to-Air setup may provide a solution for red-light (wavelength of about 620 nm to 700 nm) selective scattering. And in terms of forward scattering, for both setups of Air-to-TiO₂ and TiO₂-to-Air, differences between red-light and blue-light scattering magnitudes are not significant, so both setups of Air-to-TiO₂ and TiO₂-to-Air may achieve simultaneous selective scattering of blue- and red-light. Generally, metallic nanoparticle's resonance peak positions as well as magnitude of light scattering are also affected by metallic nanoparticle's size.[16-17] With the above considerations, forward (Figure 4 (a), (b)) and backward (Figure 5 (a), (b)) scattering cross sections are calculated for TiO₂ substrate/Ag nanocubes of several sizes (edge length = 90, 100, 110, 120, 130 nm), for both setups of Air-to-TiO₂ and TiO₂-to-Air, to search for the best solutions for selective scattering of blue- and red-light.

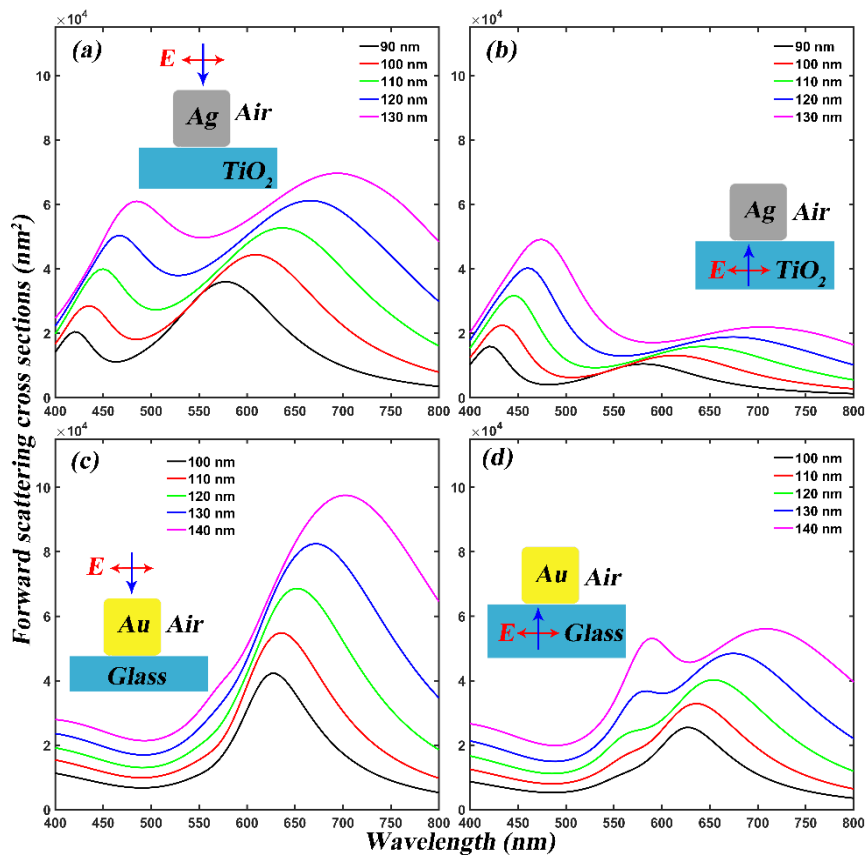

***Figure 4.*** *Simulated forward scattering cross sections for TiO₂-substrate/Ag nanocubes are shown in (a) and (b), and for glass-substrate/Au nanocubes are shown in (c) and (d). Incident direction and polarization of light for each sub-figure are illustrated by its corresponding inset. Figure legends denote the edge lengths of nanocubes. For all simulations shown in this figure, a 2-nm gap between the bottom of nanocube and the dielectric substrate is assumed to simulate the protective reagent in a real situation,[2] such as PVP (Polyvinyl Pyrrolidone). The scale of the vertical axis is the same for all sub-figures for easy comparison.*

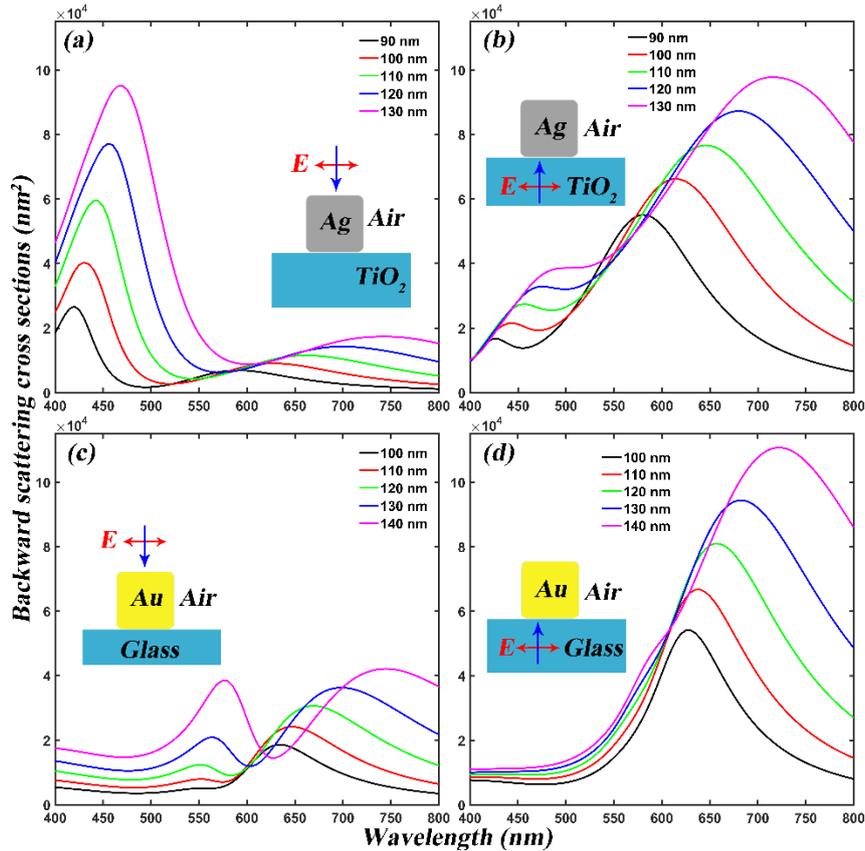

***Figure 5.*** *Simulated backward scattering cross sections for TiO₂-substrate/Ag nanocubes are shown in (a) and (b), and for glass-substrate/Au nanocubes are shown in (c) and (d). Incident direction and polarization of light for each sub-figure are illustrated by its corresponding inset. Figure legends denote the edge lengths of nanocubes. For all simulations shown in this figure, a 2-nm gap between the bottom of nanocube and the dielectric substrate is assumed to simulate the protective reagent in a real situation,[2] such as PVP (Polyvinyl Pyrrolidone). The scale of the vertical axis is the same for all sub-figures for easy comparison.*

From Figure 4 (a) and (b) and Figure 5 (a) and (b), it is observed that the scattering peaks intended for red-light selective scattering at longer wavelengths are broader than those intended for blue-light selective scattering at shorter wavelengths, but narrow scattering peak width is more desirable. To search for red-light scattering peak with narrow width, forward (Figure 4 (c), (d)) and backward (Figure 5 (c), (d)) scattering cross sections are calculated for glass substrate/Au nanocubes of several sizes (edge length = 100, 110, 120, 130, 140 nm), for both setups of Air-to-Glass (light is incident from air to glass) and Glass-to-Air (light is incident from glass to air). The reason to investigate scattering properties of Au nanocubes is that Au has a high quality factor (defined by $-Re[\varepsilon]/Im[\varepsilon]$, where $\varepsilon$ is metal's dielectric function, and the higher this value, the stronger the resonance strength) around the wavelength range of red light,[18] which suggests Au nanocubes may generate sharp and strong scattering peaks for red-light. And the reason for choosing glass rather than TiO$_2$ as the dielectric substrate for Au nanocubes is given in the following. The simulation results obtained with TiO$_2$-substrate (shown in the supplementary material) suggest that due to TiO$_2$'s high refractive index, there is a trade-off between resonance peak wavelength and magnitude of scattering, i.e., when increasing Au nanocube's size to increase magnitude of scattering cross section to a desirable value, the resonance peak wavelength is undesirably redshifted beyond 700 nm at the meantime.

A suitable scattering peak for selective scattering of blue- or red-light should satisfy two criteria: firstly, magnitude of scattering cross section around the peak wavelength should be as large as possible, since for transparent projection screen based on the dielectric substrate/metallic nanocubes structure, only one layer of metallic nanocubes can be deposited; secondly, the peak wavelength should be within the desirable wavelength ranges, i.e., for blue-light scattering, the peak wavelength should be within the range of 440 - 460 nm, and for red-light scattering, the peak wavelength should be within 620 - 700 nm. Therefore, according to these criteria and based on the observations from Figures 4 and 5, suitable setups for selective scattering of blue- and red-light can be summarized below:

- Forward scattering of blue-light: 120-nm Ag nanocube with TiO$_2$-to-Air (Figure 4 (b) blue curve).
- Forward scattering of red-light: 120-nm Ag nanocube with Air-to-TiO$_2$ (Figure 4 (a) blue curve), and 130-nm Au nanocube with Air-to-Glass (Figure 4 (c) blue curve).
- Backward scattering of blue-light: 120-nm Ag nanocube with Air-to-TiO$_2$ (Figure 5 (a) blue curve).
- Backward scattering of red-light: 120-nm Ag nanocube with TiO$_2$-to-Air (Figure 5 (b) blue curve), and 130-nm Au nanocube with Glass-to-Air (Figure 5 (d) blue curve).

So, the following two combinations can achieve forward simultaneous selective scattering of blue- and red-light: 120-nm Ag nanocube with TiO$_2$-to-Air (Figure 4 (b) blue curve) + 120-nm Ag nanocube with Air-to-TiO$_2$ (Figure 4 (a) blue curve), and 120-nm Ag nanocube with TiO$_2$-to-Air (Figure 4 (b) blue curve) + 130-nm Au nanocube with Air-to-Glass (Figure 4 (c) blue curve). Similarly, to achieve backward simultaneous selective scattering of blue- and red-light, there are also two combinations: 120-nm Ag nanocube with Air-to-TiO$_2$ (Figure 5 (a) blue curve) + 120-nm Ag nanocube with TiO$_2$-to-Air (Figure 5 (b) blue curve), and 120-nm Ag nanocube with Air-to-TiO$_2$ (Figure 5 (a) blue curve) + 130-nm Au nanocube with Glass-to-Air (Figure 5 (d) blue curve). And it is noted that the combination (120-nm Ag nanocube with Air-to-TiO₂) + (120-nm Ag nanocube with TiO₂-to-Air) appears both in forward and backward scattering, which indicates that this combination can achieve both forward and backward selective scattering of blue- and red-light. Therefore, there are only three combinations for simultaneous scattering of blue- and red-light, and are summarized below:

1. Forward scattering: (120-nm Ag nanocube with TiO₂-to-Air) + (130-nm Au nanocube with Air-to-Glass)
2. Backward scattering: (120-nm Ag nanocube with Air-to-TiO₂) + (130-nm Au nanocube with Glass-to-Air)
3. Forward and backward scattering: (120-nm Ag nanocube with Air-to-TiO₂) + (120-nm Ag nanocube with TiO₂-to-Air)

The schematics of the three combinations are shown in insets of Figure 6, with TiO₂ as coating layer(s) on glass, and glass as the substrate. Interface between two different media reflects light, thus the amount of light reaching the back side (as opposed to the side where incident light is injected) of the TiO₂-coated glass slip is reduced, which decreases scattering of metallic nanocubes placed on this side. So, it is important to reduce light reflection by properly selecting thickness of TiO₂ layer(s) coated on glass. For the first combination, it is desirable to reduce reflection of red-light since Au nanocube placed on the back side is responsible for red-light backward scattering. Reduction of red-light reflection can be achieved by a destructive interference between the light reflected at the air/TiO₂ interface and the light reflected at the TiO₂/glass interface. Light reflection at the air/TiO₂ interface has a phase-shift of $180°$, while light reflection at the TiO₂/glass (TiO₂/air) has no phase-shift, so a destructive interference between these two reflections is realized by choosing the TiO₂ thin film's thickness to be $\lambda_{red}/(2n)$, where $\lambda_{red}$ is the red light's wavelength in vacuum and is chosen to be 650 nm in our work, and $n$ is the refractive index of the TiO₂ thin film which is measured to be 2.3. Thus, thickness of the TiO₂ thin film for the first combination is decided to be 137 nm. Similarly, for the second combination, since Ag nanocube placed on the back side is responsible for blue-light forward scattering, reduction of blue-light reflection is desired and thickness of the TiO₂ thin film of this combination is decided to be 100 nm. For the third combination, Ag nanocube placed on the back side is responsible for both blue- and red-light scattering, so to achieve a balance, the wavelength at which the light's reflection to be minimized is selected to be 550 nm, leading to a thickness of 120 nm for TiO₂ thin film of this combination.

For the first combination, backward scattering from Ag nanocube is combined with that from Au nanocube, with consideration of TiO₂ thin film's thickness and light reflections at all interfaces, and is shown as an overall backward scattering in Figure 6 (a). Similarly, combined overall forward scattering for the second combination is shown in Figure 6 (b), and both combined overall forward and backward scattering spectrums for the third combination are shown in Figure 6 (c).

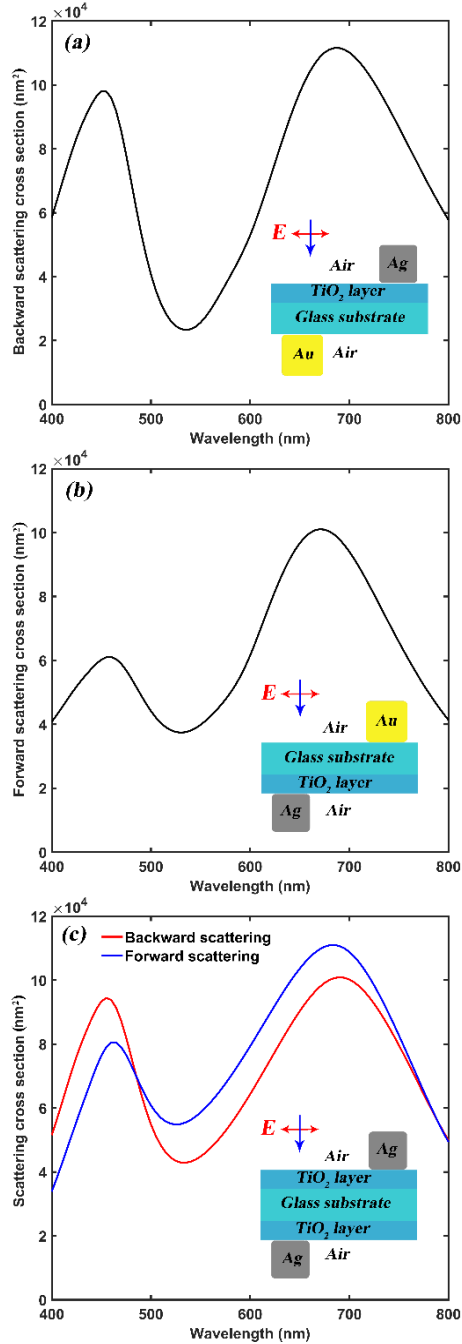

**Figure 6.** *Combined overall forward or backward scattering cross sections for the three combinations discussed in text, with their schematics shown by insets, and effect of TiO₂ thin film's thickness and light reflections at all interfaces are considered for all three combinations. (a) Overall backward scattering for the first combination combined from that of Ag nanocube and Au nanocube, thickness of the TiO₂ thin film is 137 nm. (b) Overall forward scattering for the second combination combined from that of Ag nanocube and Au nanocube, thickness of the TiO₂ thin film is 100 nm. (c) Overall forward scattering as well as backward scattering for the third combination combined from that of Ag nanocube at the foreside and that of Ag nanocube at the backside, thickness of both the two TiO₂ thin films is 120 nm. In (a), (b) and (c), Ag nanocube's edge length is 120 nm. In (a) and (b), Au nanocube's edge length is 130*

*nm. For all simulations shown in this figure, a 2-nm gap between the bottom of nanocube and the dielectric substrate is assumed to simulate protective reagent in a real situation,[2] such as PVP (Polyvinyl Pyrrolidone).*

From Figure 6, it is observed that all three combinations achieve selective enhancement of blue- and red-light scattering, either in forward or backward or both directions. And given a scattering cross section of about 50000 nm$^2$ at peak wavelength for a single nanoparticle is able to generate a clear projected image with proper areal nanoparticle density as demonstrated in previous work,[2] all three combinations as shown in Figure 6 have enough magnitudes of scattering and are expected to generate clear projected image with suitable areal concentration of nanocubes. In Figure 6 (b), the relative weaker peak around 455 nm for blue-light forward scattering can be strengthened by increasing areal concentration of Ag nanocubes deposited on the backside of the TiO$_2$-coated glass slide relative to that of Au nanocubes deposited on the foreside in a real application. For each combination shown in Figure 6, angular distribution of scattering is plotted for both peaks and are shown in Figure 7.

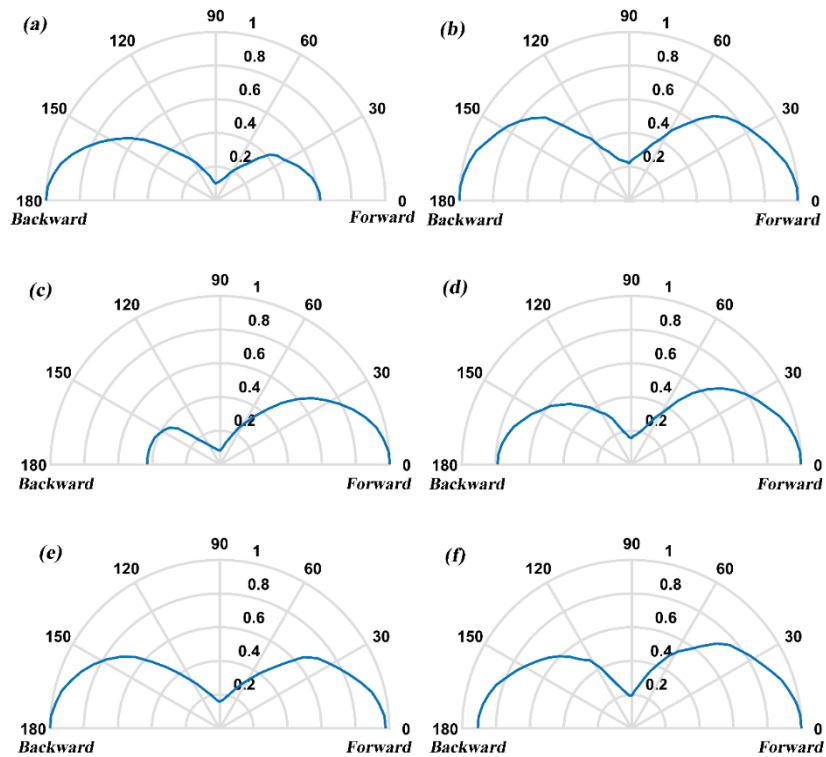

**Figure 7** *Normalized scattering intensity |E|$^2$ plotted versus scattering angle for the three combinations shown in Figure 6. (a) Plotting for the first combination at the wavelength of 455 nm. (b) Plotting for the first combination at the wavelength of 680 nm. (c) Plotting for the second combination at the wavelength of 460 nm. (d) Plotting for the second combination at the wavelength of 680 nm. (e) Plotting for the third combination at the wavelength of 460 nm. (f) Plotting for the third combination at the wavelength of 685 nm. All scattering intensities shown in this figure consist of*

contributions from both the nanocubes placed on the foreside and backside of the TiO₂-coated glass slide. Incident light has equal components for s- and p- polarisations with respect to the scattering plane. Each sub-figure's angular distribution of scattering intensity is normalized to its largest value.

From Figure 7, it is observed that the angular distribution of scattering is relatively uniform for both peaks in backward direction (90° to 180°) for the first combination, in forward direction (0° to 90°) for the second combination, and in both forward and backward directions (0° to 180°) for the third combination, which indicate projected image should have large viewing-angle range for all the three combinations shown in Figure 6.

**Conclusion**

In this work, through numerical simulations, forward and backward scattering properties of Ag and Au nanocubes placed on dielectric substrate are investigated, with effect of nanocube's size and both setups of Air-to-Dielectric (light is incident from air to dielectric) and Dielectric-to-Air (light is incident from dielectric to air) taken into account. Based on the forward and backward scattering properties, three designs (the three combinations shown in Figure 6) are proposed, two of which achieve selective scattering of blue- and red-light only in either forward or backward direction, and the other one achieves selective scattering of blue- and red-light both in forward and backward directions, all with reasonable large magnitudes of scattering strength. Simulated angular distribution of scattering suggests that the viewing-angle range is large for all the three designs.

**Acknowledgement**

This work was financially supported by the National Research Foundation of Singapore (Program Grant No. NRF-CRP13-2014-02).